\documentclass[twocolumn,preprintnumbers,amsmath,amssymb,aps]{revtex4}
 \normalsize

\def\lsim{\raise0.3ex\hbox{$\;<$\kern-0.75em\raise-1.1ex\hbox{$\sim\;$}}}
\def\gsim{\raise0.3ex\hbox{$\;>$\kern-0.75em\raise-1.1ex\hbox{$\sim\;$}}}

\def\2tvec#1#2{ \left( \begin{array}{c}
#1  \\
#2  \\
\end{array} \right)}%
\def\mat2#1#2#3#4{ \left( \begin{array}{cc}
#1 & #2 \\
#3 & #4 \\
\end{array} \right) }%
\def\Mat3#1#2#3#4#5#6#7#8#9{ \left( \begin{array}{ccc}tri-bimaximal
#1 & #2 & #3 \\
#4 & #5 & #6 \\
#7 & #8 & #9 \\
\end{array} \right) } %
\def\Mat3#1#2#3#4#5#6#7#8#9{ \left( \begin{array}{ccc}
#1 & #2 & #3 \\
#4 & #5 & #6 \\
#7 & #8 & #9 \\
\end{array} \right) }

\def\3tvec#1#2#3{ \left( \begin{array}{c}
#1  \\
#2  \\
#3  \\
\end{array} \right)}

\def\4tvec#1#2#3#4{ \left( \begin{array}{c}
#1  \\
#2  \\
#3  \\
#4  \\
\end{array} \right)}

\def\hbar{\hspace{1mm}\bar{}\hspace{-1mm}h}

\newcommand{\balg}{\begin{align}}
\def\bea{\begin{eqnarray}} \def\eea{\end{eqnarray}}
\newcommand{\be}{\begin{eqnarray}} \newcommand{\ee}{\end{eqnarray}}
\usepackage{feynmp} \usepackage{slashed} \usepackage{latexsym}
\usepackage{graphicx} \usepackage{verbatim}
\usepackage[normalem]{ulem} 

\begin{document}

\title{
Neutrino Masses and Deviation from Tri-bimaximal mixing in $\Delta(27)$ model with Inverse Seesaw Mechanism }%

%
\author{M. Abbas$^{1,2}$, S. Khalil$^{2,3}$, A. Rashed$^{1,2}$ and A. Sil$^{4}$  }
\vspace*{0.2cm}
\affiliation{$^1$Department of Physics, Faculty of Science,  Ain Shams University, Cairo, 11566, Egypt.\\
$^2$Center for Fundamental Physics, Zewail City of Science and Technology, Giza 12588, Egypt.\\
$^3$Department of Mathematics, Faculty of Science,  Ain Shams University, Cairo, 11566, Egypt. \\
$^4$Indian Institute of Technology Guwahati, 781039 Assam, India.}
%
%

\begin{abstract}
\noindent We propose a scheme, based on $\Delta(27)$ flavor symmetry and supplemented by other discrete symmetries and 
inverse seesaw mechanism, where both the light neutrino masses and the deviation from tri-bimaximal mixing matrix can be linked 
to the source of lepton number violation. The hierarchies of the charged leptons are explained. We find that the quark masses 
including their hierarchies and the mixing can also be constructed in a similar way. 

\end{abstract}

\maketitle

The convincing evidence of small but non-vanishing neutrino masses calls for
an explanation from a naturalness point of view. It actually points to the
existence of new physics beyond the electroweak scale ($v$). There exist
several scenarios to explain this smallness of neutrino masses. Among them,
perhaps the most well-studied one is the conventional type-I seesaw mechanism \cite{T1}. 
In this mechanism, the smallness of neutrino mass ($m_{\nu}$) can be obtained
in an economic way at the expense of introducing heavy right handed (RH)
neutrinos ($\nu_R$). For values of Yukawa couplings involved ($Y_{\nu}$) of
order unity, the mass scale of $\nu_R$ ($M_R$) turns out to be near the
grand unified scale or so through the relation $m_{\nu} = - m_D {M_R}^{-1} m_D^T$,
where $m_D = Y_{\nu} v$. Although interesting, such a large scale is beyond the 
experimental reach.

In this regard, the inverse seesaw mechanism \cite{Wyler:1982dd, Mohapatra:1986bd, 
Ma:1987zm} offers an interesting resolution through a double suppression by the 
new physics scale $M$ through $m_{\nu} =m_D {M}^{-1} \mu {M^T}^{-1}m_D^T$. 
With a small mass scale  $\mu$ (of order KeV to few hundred MeV), a relatively low new 
physics scale (accessible to LHC) associated with $M$ results. However the main caveat of this scenario 
is to understand the smallness associated with $\mu$ or in other words, how it is generated. 
Note that in case of type-I seesaw, the lepton number violation (LNV) happens
through the majorana mass term of the  RH neutrinos, which is quite large. Contrary to this, 
in case of inverse-seesaw, it happens via the $\mu$ term which is a tiny scale while 
compared to the electroweak scale. As the lepton number is only an approximate symmetry 
of nature, it would be more natural to break it by a small amount rather than by a mass 
term like $M_R$, which is very large. It can also be argued from the sense of 't Hooft \cite{'t Hooft},  
just because in the limit $\mu$ tends to zero, the $m_{\nu}$ goes to zero and LNV vanishes so that 
the symmetry is enhanced.

In this letter, we explain the desired smallness of $\mu$-term in a flavor symmetric framework.
We consider the presence of a $\Delta(27)$ flavor symmetry which is supplemented by 
additional $Z_4 \times Z_3$ discrete groups. The structure guarantees the non-appearance 
of the $\mu$-term in the tree level Lagrangian. In fact, it allows the $\mu$-term to be generated 
only through a significantly higher dimensional operator and thereby suppressing the corresponding 
interaction by some nonzero powers of the cut-off scale ($\Lambda$) of the theory.
There are flavon fields, whose vacuum expectation values (vev) would break the flavor 
symmetry and thereby generates a specific structure of $\mu$ and other mass matrices 
like neutrino Dirac mass matrix ($m_D$), charged leptons etc. We will elaborate more on this 
as we proceed.  In addition, we assume a 2-3 flavor symmetry as an additional symmetry of the 
Lagrangian (particularly for the lepton sector). The only place where this 2-3 symmetry will be 
broken is in the vev alignment of a single flavon field ($\sigma$) responsible for generating the $\mu$ term. 
The vev of all other flavons respect the 2-3 symmetry. So,  in a way our framework suggests a unified 
source (through $\mu$ term only) of breaking the 2-3 symmetry and lepton number violation.
It is known \cite{Abbas:2014ewa} that a breaking of 2-3 symmetry may indicate a deviation from tri-bimaximal mixing 
in the neutrino sector. Therefore in this work, we argue that the specific structure obtained for 
$\mu$ not only can explain the small masses of light neutrinos, but also accounts for the deviation 
from an exact tri-bimaximal mixing by having a nonzero $\theta_{13}$ at the same time. 

In realizing the above goal, the fermion field content of the Standard Model (SM) is extended by 
adding three right handed neutrinos $\nu_{R_i}$ (for $i= 1, ~2, ~3$), three SM gauge singlet 
fermions ${S_i}$ which have lepton number opposite to that of the $\nu_{R_i}$. In addition, 
the scalar sector is extended by adding a set of flavons that break the flavor symmetry around 
few TeV scale or more.  In \cite{Abbas:2014ewa}, it was emphasized that both quark and lepton 
masses and also their mixing angles can be simultaneously accommodated in a framework of $\Delta (27)$ 
based on the type-I seesaw mechanism. Here also  we have considered the quark sector. We have 
constructed both the up-type and down-type quark mass matrices so that an acceptable 
Cabbibo-Kobayashi-Maskawa mixing matrix ($V_{CKM}$) can be obtained. In realizing the $V_{CKM}$, 
we have employed only one additional flavon apart from those are already involved in the lepton sector.

It turns out that the  $\Delta(27)$ symmetry alone is not enough to restrict all allowed 
Yukawa interactions that could lead to consistent mass matrices, additional
$Z_4 \times Z_3$ symmetries are also imposed as mentioned in Table \ref{table1}.
$\chi$ and $\sigma$ are the $\Delta$(27) triplet flavons  and $\eta_1$ is the singlet 
flavon field. 
\begin{table}[t]%
\begin{tabular}{|c|c|c|c|c|c|c|c|c|c|c|c|c|}
  \hline
   Fields  & ~ $\ell_i$ ~ & ~ $E_{Ri}$  ~ & ~ $\nu_{R_i}$ ~ & ~ $S_{i}$ ~ & ~ $H$ ~ & ~ $\chi$ ~ & ~ $\sigma$~& ~ $\eta_{1}$~ & ~$\eta_{2}$  \\
  \hline $\Delta (27)$&3&$3$ & $1_i$ & $3 $ & $1_1$ & $3$ & $3$ & $1_1$ &  $1_2$ \\
   \hline $Z_4$ & 1 &-1& i& -1& 1&-i& -i & $-1$ & $ 1$\\
  \hline $Z_3$& 1 & $\omega^2$& $\omega$& $\omega$& 1&$\omega^2$& $\omega^2$ & $\omega$i & $ 1$\\
   \hline $L$& 1 & 1& 1& -1& 0&0& $1/2$ & $0$ & $0$\\
  \hline
\end{tabular} \centering \caption{Field transformations under
$\Delta (27)$ , $Z_4$ and $Z_3$.  Here $i = 1,2, 3$ refer to the generation indices. $L$ 
stands for the lepton number.}
\label{table1} \end{table}
The relevant terms in the Lagrangian for neutrino mass which are consistent with the symmetries
considered above are given by%
\bea%
{\cal L}=\frac{h^k_{ij}}{\Lambda} \bar{\ell_i}\tilde{H}\nu_{R_j}
\chi_k+\frac{f^k_{ij}}{\Lambda^5}\bar{S}_i^c~S_j\sigma_k^4 \eta_1^2+g^k_{ij}
\bar{\nu}_{R_i}^c
S_{j}\chi_k^\dagger.%
\label{lagranigan}
\eea%
Here $i,j$ are the flavor indices while $k$ refers to the $k^{th}$ component of $\Delta$(27) triplet flavon 
field only. Once the $\chi$ and $\sigma$ fields obtain vevs, $\Delta(27)$ symmetry is broken. This 
breaking leads to specific flavor structures for the heavy mass matrix $M =g \langle \chi \rangle $ 
and the $\mu$-term $\mu =  f \langle \sigma \rangle^4 \langle \eta_1 \rangle^2/\Lambda^5$, 
where the flavor indices are suppressed for simplicity. Each of these are $3 \times 3$ mass matrices. 
Note that the second term in Eq.(\ref{lagranigan}) is highly suppressed, thereby 
has the potential to explain the smallness associated with the $\mu$-term in an inverse seesaw 
framework. The electroweak symmetry breaking thereafter results in the Dirac neutrino mass term: 
$m_D =  h \langle \chi \rangle \langle H \rangle/\Lambda$. Hence the neutrino mass matrix, 
in the basis $\{\nu_L^c, \nu_R, S\}$, is given by%
\bea%
M_\nu=\left(
    \begin{array}{ccc}
      0 & m_D & 0 \\
      m_D^T & 0 & M \\
      0 & M^T & \mu \\
    \end{array}
  \right),%
  \eea%
which is a $9 \times 9$ matrix. It can be noted a term involving the operator 
$\ell H \ell H $ (which could contribute to the 11 block of $M_{\nu}$) is absent in our set-up upto dimension-9 with the 
symmetries we considered. 

At this moment, a discussion about the 2-3 symmetry over the flavor or generation indices 
would be pertinent here. It is well known \cite{Barger:2012fx}, \cite{Lam:2005va} that 
with a 2-3 symmetry,  tri-bimaximal mixing can be achieved at the zeroth order. We have 
also considered this 2-3 symmetry for the lepton sector on top of the discrete symmetries 
listed in Table \ref{table1}. Therefore particles transform as  $f_1 \leftrightarrow f_1$ and $f_2 \leftrightarrow 
f_3$ with $f_i$ stands for $\ell_i$, $\chi_i$, $\sigma_i$, $S_i$ and $\nu_{R_i}$. This 
choice further simplifies the structure of $m_D$, $M$ and $\mu$ which will be discussed 
later. Regarding the vevs of the triplets involved in the model, we have chosen a specific 
alignment as 
\begin{equation}
\langle\chi\rangle=(v_{\chi}, v_{\chi}, v_{\chi}), ~~~~\langle\sigma\rangle=(0, v_{\sigma}, 0).
\label{vev-align}
\end{equation}
Our choice is governed by the fact that it breaks both the $\Delta(27) (\equiv Z^{'}_3 \times Z^{''}_3 \times Z^{'''}_3)$
and the $2-3$ symmetry as well. One can easily see that $\langle\chi\rangle$ breaks
$Z^{'}_3 \times Z^{''}_3$ subset of $\Delta(27)$, while  $\langle \sigma \rangle$ breaks  
$Z^{'''}_3 \times 2-3$ symmetry. We will show that this small breaking (as $\mu$ is small) 
of 2-3 symmetry allows to have non-zero $\theta_{13}$ in our framework. 

In order to analyze the vevs of the triplets $\sigma$ and $\chi$, we consider the most general scalar potential  
invariant under the symmetries considered, which is given by
\bea
V \!&\!=\!&\! m_\sigma^2 (\sigma^\dagger \sigma)+m_\chi^2 (\chi^\dagger \chi)+\lambda_\sigma 
(\sigma^\dagger \sigma) (\sigma^\dagger \sigma) \nonumber\\
\!&\!+\!&\! \lambda_\chi (\chi^\dagger \chi) (\chi^\dagger \chi) + \kappa \left[ (\sigma^\dagger \sigma) 
(\chi^\dagger \chi) +  (\sigma^\dagger \chi) (\chi^\dagger \sigma)\right], ~~
\eea
where $\kappa > -2 \sqrt{ \lambda_{\sigma} \lambda_{\chi}}$ and $\lambda_{\sigma} , \lambda_{\chi} \geq 0$, 
so that the potential is bounded from below. For simplicity we assume universal coupling for the last two terms 
in the above potential. In general this potential contains several free parameters (masses and couplings).
These  plenty of free parameters allow all type of patterns of non-zero vevs of $\sigma$ and $\chi$.
Once we restrict ourselves with the particular vev alignments of $\chi$ and $\sigma$ as given in Eq.(\ref{vev-align}), 
the following non-zero vevs are obtained:
\bea
v^2_{\chi} = -\frac{1}{4} \frac{\left[ 2 \kappa m_{\sigma}^2 + 3 \lambda_{\sigma} m_{\chi}^2 \right ]}{\left [4\kappa^2 +9 \lambda_{\chi} \lambda_{\sigma}\right ]},~~
v^2_{\sigma} = \frac{1}{2} \frac{\left[ 2\kappa m_{\chi}^2 - 3 \lambda_{\chi} m_{\sigma}^2\right ]}{\left[ 4\kappa^2 + 9 \lambda_{\chi} \lambda_{\sigma}\right ]}.~
\eea
We assume that the vev's of $\chi $ and $\sigma$ are all of the same order and satisfy the
following relation $\frac{v_{\chi}}{\Lambda} \sim \frac{v_{\sigma}}{\Lambda} = \frac{u}{\Lambda} \sim{\cal O}(\lambda_C^2)$ where $\lambda_C$ is the Cabibbo
angle, i.e. $\lambda_C \sim 0.22$.

We now analyze the detailed flavor structure of the block matrices involved in $M_{\nu}$. 
From Eq. (\ref{lagranigan}), one finds that the Dirac neutrino mass
matrix $m_D$ can be constructed from the following invariants terms:%
\bea%
 && \frac{1}{\Lambda}h_1 (\bar\ell_1 \chi_1+\bar\ell_2
\chi_2+\bar\ell_3 \chi_3) \nu_{R_1}\tilde{H},\nonumber\\
&& \frac{1}{\Lambda}h_2 (\bar\ell_1 \chi_1+\omega^2 \bar\ell_2
\chi_2+\omega \bar\ell_3
\chi_3) \nu_{R_2}\tilde{H},\nonumber\\
&& \frac{1}{\Lambda}h_3 (\bar\ell_1 \chi_1+\omega \bar\ell_2
\chi_2+\omega^2 \bar\ell_3 \chi_3) \nu_{R_3}\tilde{H}, \nonumber
\eea%
where $h^k_{ij}$ is written as $h_k$ only.
Thus, after the flavor symmetry breaking, the following Dirac neutrino
mass matrix is obtained %
\bea%
m_D=\frac{u \langle H\rangle}{\Lambda}\left(
    \begin{array}{ccc}
      h_1  & h_2  & h_2  \\
      h_1  & \omega^2 h_2 & \omega h_2  \\
      h_1 & \omega h_2  & \omega^2 h_2 \\
    \end{array}
  \right),%
  \eea%
where $h_2=h_3$ is considered due to the presence of $2-3$  symmetry in the Lagrangian.
Similarly, the matrix $M$ takes the  form%
\bea%
M=u \left(
    \begin{array}{ccc}
      g_1  & g_1  & g_1 \\
      g_2  & \omega^2 g_2 & \omega g_2  \\
      g_2  & \omega g_2  & \omega^2 g_2 \\
    \end{array}
  \right).%
  \eea%
Finally, the mass matrix $\mu$ is given by  %
\bea%
\mu=\frac{u^6}{\Lambda^5}\left(
      \begin{array}{ccc}
        0 &  0 &  f_2  \\
         0 &  f_1  &  0 \\
         f_2  &  0 &  0 \\
      \end{array}
    \right),
    \eea%
where $\langle \eta_1 \rangle$ is assumed to be of order $u$. Note that the $2-3$ 
symmetry is violated only by  $\langle \sigma \rangle$ , which would be the source of 
deviation from tribimaximal mixing pattern as well in the lepton sector.

The diagonalization of $M_\nu$ mass matrix leads to the following light and heavy neutrino masses respectively:
\begin{equation}
M_{\nu_l}= m_D~(M~\mu^{-1}~M^T)^{-1}~m_D^T,
\label{mnul}
\end{equation}
\begin{equation}
M_{\nu_H}= M_{\nu_H'}=\sqrt{M^2 + m_D^2}.
\label{mnu}
\end{equation}
It is now clear that the light neutrino masses can be of order eV, with a TeV scale $M$ provided
$\mu \ll m_D, M$. We will show that with the charge assignments we have considered, the charged 
lepton mass matrix comes out to be a diagonal one. 
Therefore the light neutrino mass matrix  $M_{\nu}$ must be diagonalized by the physical neutrino mixing
matrix $U_{PMNS}$, {\it i.e.},
$ U_{PMNS}^T \,M_{\nu_l} \,U_{PMNS} = {\rm diag} (m_1, m_2, m_3).$
Since $\mu$ matrix is generated by $\langle \sigma \rangle$, which violates the 2-3
symmetry while $m_D$ and $M$ are 2-3 symmetric,
the deviation of $U_{PMNS}$ from tri-bimaximal mixing is proportional to the size of $\mu$,
which is quite suppressed. A discussion regarding the lepton number is appropriate here. We have
considered the entire Lagrangian to respect the lepton number. However, $\sigma$ having a
nonzero lepton number, while gets a vev, the lepton number is broken and in turn generation of neutrino masses
results.

We define the parameters which
characterize the deviation of mixing angles from
the tri-bimaximal  values as %
\bea %
D_{12}\equiv \frac{1}{3}-s_{12}^2,~~ D_{23} \equiv
\frac{1}{2}-s_{23}^2, ~~ D_{13}\equiv s_{13},
\label{TBMdeviation} %
\eea%
where $s_{ij}\equiv\sin\theta_{ij}$.  The tri-bimaximal mixing matrix corresponds to neutrino mass matrix that satisfies the following three conditions \cite{Abbas:2010jw}:%
$(M_{\nu_l})_{12}=(M_{\nu_l})_{13}$, $(M_{\nu_l})_{22}=(M_{\nu_l})_{33}$, and $(M_{\nu_l})_{11}+(M_{\nu_l})_{12}=(M_{\nu_l})_{22}+(M_{\nu_l})_{23}$.
Therefore, the deviation of tri-bimaximal mixing matrix can be written
in terms of the deviation parameters $D_{23}$ and $s_{13}$ as follows: %
\begin{widetext}
\bea
\Delta_1&=&(M_{\nu_l})_{12}-(M_{\nu_l})_{13}=\frac{\sqrt{2}}{3}\left[ (2 m_1+m_2)e^{2i\delta}-3m_3 \right] s_{13}e^{-i\delta}+\frac{2}{3}(m_2-m_1)D_{23},\nonumber\\%
\Delta_2&=&(M_{\nu_l})_{22}-(M_{\nu_l})_{33}=\frac{2\sqrt{2}}{3}(m_2-m_1)s_{13}e^{i\delta}+\frac{1}{3}(m_1+2m_2-3m_3)D_{23},
\label{Delta1}
\eea%
 \end{widetext}
where $m_i$ is the physical neutrino mass and $\delta$ is the
leptonic Dirac phase. However, for simplicity, we set the Dirac phase to be zero. 
In our model the deviations from TBM
conditions can give constraints on our parameters (couplings and
VEVs) in order to get the correct mixing angles and desired scenario
of mass spectra. From Eq.~(\ref{mnul}), one can write the equations above in terms of the model parameters\\[10mm]

\bea%
\Delta_1&=&-\frac{h_2 \langle H \rangle^2 (-f_1 g_2 h_1+f_2 g_2
h_1+f_1g_1h_2+2
f_2g_1h_2)u^4}{3 g_1g_2^2\Lambda^5}, \nonumber\\
\Delta_2&=&\frac{h_2 \langle H \rangle^2 (2f_1 g_2 h_1-2f_2 g_2
h_1+f_1g_1h_2+2
f_2g_1h_2)u^4}{3 g_1g_2^2\Lambda^5}.\label{Delta2}\nonumber\\
\eea%
From Eqs. (\ref{Delta1}) and (\ref{Delta2}) we can calculate the deviation from TBM parameters %

\begin{widetext}
\bea%
s_{13}&=&-\frac{\langle H\rangle^2 h_2 \left[ (f_1-f_2) g_2 h_1 (5 m_1 - 2 m_2 - 3 m_3) + g_1 h_2 (f_1+2f_2) (m_1 - 4 m_2 + 3 m_3) \right] u^4}{ \sqrt{2} g_1 g_2^2 (2 m_1^2 - 13 m_1 m_2 + 2 m_2^2 + 9 (m_1 + m_2) m_3 - 9 m_3^2) \Lambda^5},\nonumber\\
D_{23}&=&\frac{3 \langle H\rangle^2 h_2 \left[ 2 (f_2-f_1) g_2 h_1 (m_1 - m_3) +(f_1 + 2 f_2) g_1 h_2 (-m_2 + m_3) \right] u^4}{g_1 g_2^2 (2 m_1^2 - 13 m_1 m_2 + 2 m_2^2 + 9 (m_1 + m_2) m_3 - 9 m_3^2)\Lambda^5}.\eea
\end{widetext}
From these expressions, one can easily see that both the $\sin\theta_{13}$ and the $D_{23}$ are proportional 
to $\mu/u^2$. Therefore in the limit as $\mu$ tends to zero, $s_{13}, D_{23}$ as well as the neutrino mass 
vanish. 

\begin{figure}[htb!]
\centering
\includegraphics[width=6.5cm,height=8.5cm,angle=-90]{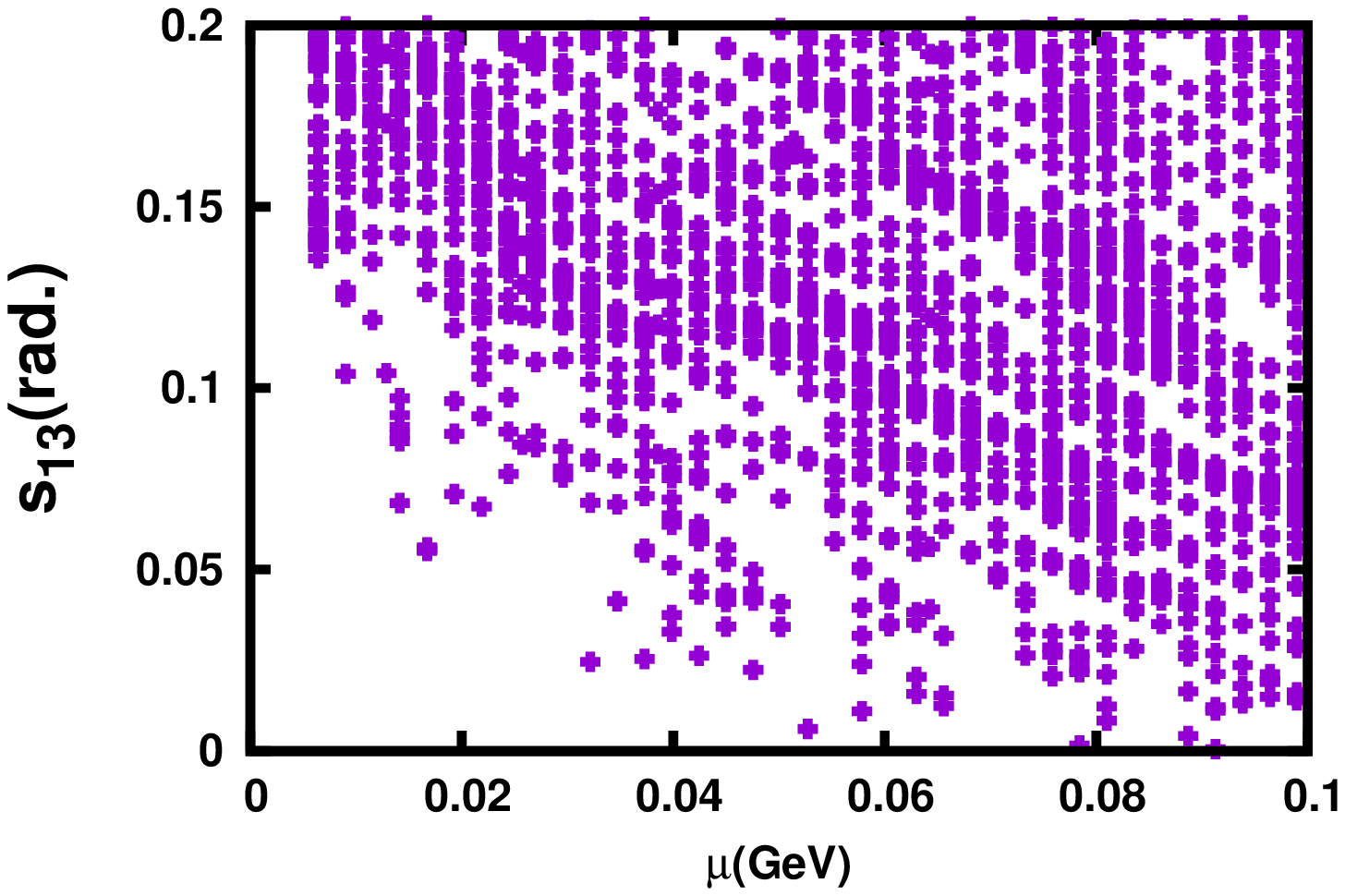}

\includegraphics[width=6.5cm,height=8.5cm,angle=-90]{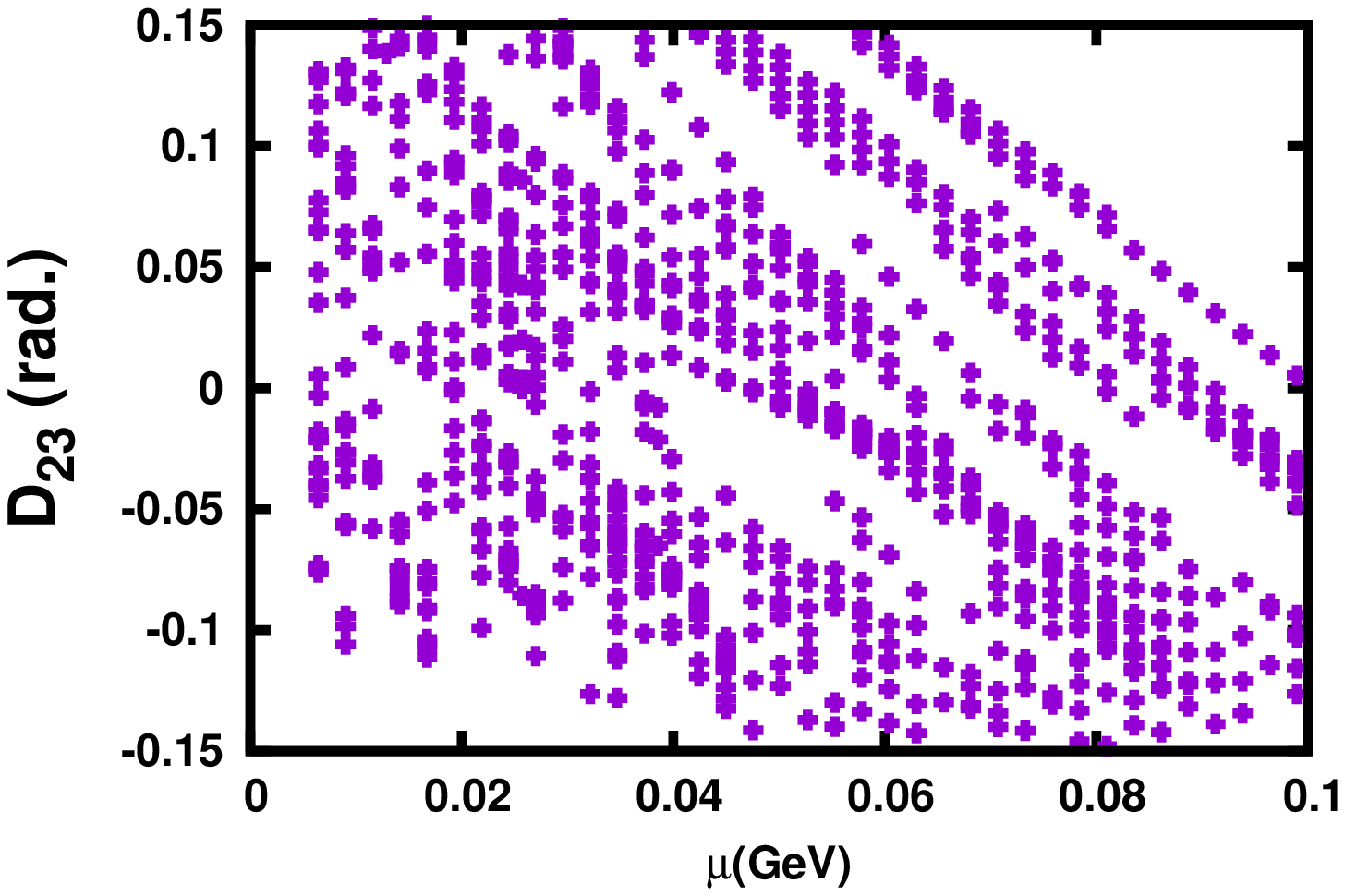}
\caption{The variation of the parameters $s_{13}$ and $D_{23}$ with respect to the 
mass scale associated with the $\mu$ term are displayed.}
\label{fig1b}
\end{figure}

As shown in Fig.~\ref{fig1b}, we can tune the involved parameters to get $s_{13}$ 
and $D_{23}$ within their experimental limits:  $0.137\leq s_{13}\leq 0.158$ and 
$-0.144\leq D_{23}\leq 0.115$ \cite{Gonzalez-Garcia:2014bfa}. Here the cut-off 
scale is fixed at $\Lambda  = 10^7$ GeV and the vev $u \simeq {\cal O}(100)$ TeV.  
The couplings involved in the $m_D, ~M$, and $\mu$ matrices, although tuned, are considered 
to be of order unity. The allowed regions are consistent with the neutrino 
oscillation parameters \cite{Gonzalez-Garcia:2014bfa} as well as satisfy the cosmological bound on 
sum of the light neutrino masses \cite{cosmo-neu}.    

The charged lepton Yukawa Lagrangian, invariant under the symmetries considered, is given by%
\bea
\!{\cal L}_l=\frac{\lambda_1}{\Lambda} \bar{\ell} H E_{R} \eta_1 + \frac{\lambda_2}{\Lambda^2} \bar{\ell} H E_{R} \eta_1 \eta_2^\dagger + \frac{\lambda_3}{\Lambda^2} \bar{\ell} H E_{R} \eta_1 \eta_2, 
\eea
where the charges of $\eta_2$ under the discrete symmetries are specified in Table \ref{table1}. 
The Lagrangian indicates that the charged lepton mass matrix is diagonal. The charged lepton masses are therefore given by,
\bea%
m_e&=&\lambda_C^2\langle H \rangle(\lambda_1+ \lambda_2 \lambda_C^2 +\lambda_3 \lambda_C^2),\nonumber\\
m_\mu&=&\lambda_C^2\langle H \rangle(\lambda_1+ \omega^2\lambda_2 \lambda_C^2+\omega \lambda_3 \lambda_C^2),\nonumber\\
m_\tau&=&\lambda_C^2 \langle H \rangle(\lambda_1+ \omega\lambda_2 \lambda_C^2+\omega^2 \lambda_3 \lambda_C^2),
\eea%
where the $\langle \eta_i \rangle/\Lambda \sim \lambda_C^2$ is considered. In general the coupling constants $\lambda_i$ are complex, so the three lepton masses with 
required hierarchy can be realized. For instance, with $\lambda_1 \simeq {\cal O}(0.1)$ and $\lambda_2 = \lambda_3^\dagger\simeq -1.12 + 1.7 i$, one finds $m_e = 0.5$ MeV, $m_\mu = 0.1$ GeV, and $m_\tau = 1.78$ GeV.

\begin{table}[th]%
\begin{tabular}{|c|c|c|c|c|c|c|c|c|c|c|c|}
  \hline
  Fields  & $\bar{Q}_1$  &  $\bar{Q}_2$  &  $\bar{Q}_3$ &  $u_{R}$ & $c_{R}$  & $t_{R}$ & $d_{R}$ & $s_{R}$ &  $b_{R}$ & $\eta_3$~ \\
  \hline $\Delta (27)$ & $1_5$ & $1_8$ & $1_2$ & $1_9$& $1_4$ & $1_3$ & $1_2$ & $1_1$ & $1_3$ & $1_4$ ~  \\
  \hline $Z_3$ & $\omega$ & $\omega^2$ & $\omega^2$ & $1$& $\omega$ & $\omega$ & $\omega^2$ & $1$ &$\omega^2$ & $\omega$~  \\
   \hline $Z_4$ & $1$ & $-1$ & $1$ & $1$& $-1$ & $1$ & $-1$ & $1$ & $-1$ &$-1$~  \\
   \hline
\end{tabular} \centering \caption{Field transformations under
$\Delta (27), Z_4$ and $Z_3$ in the quark sector.} \label{table3} \end{table}

For completeness, we briefly discuss the quark sector. In Table
\ref{table3} we present the charge assignments of up and down quarks 
under $\Delta (27)$, $Z_4$ and $Z_3$ symmetries.  We also include an
extra singlet $\eta_3$ which is necessary for building consistent
quark mass matrices. The invariant Lagrangian of up-quarks under the
above symmetries, up to operators suppressed by $1/{\Lambda^4}$, are 
given by:%
\bea%
\!\!\!\!\!\!{\cal L}_u \!\!&\!\!=\!\!&\!\! \frac{h^u_{11}}{\Lambda^2} \bar{Q}_1 \tilde {H} u_{R} \eta_1^2\eta_2^\dagger  \!+\! \frac{h^u_{12}}{\Lambda^2} \bar{Q}_1 \tilde {H} c_{R} \eta_2\eta_3 \!+\! \frac{h^u_{13}}{\Lambda^4} \bar{Q}_1 \tilde {H} t_{R} \eta_3^2\eta_2^\dagger\eta_1^2 \nonumber\\ 
&+& \frac{h^u_{21}}{\Lambda^2} \bar{Q}_2 \tilde {H} u_{R} \eta_1^2\eta_3^\dagger \!+\!  \frac{h^u_{22}}{\Lambda} \bar{Q}_2 \tilde {H} c_{R} \eta_2^\dagger \!+\! \frac{h^u_{23}}{\Lambda^2} \bar{Q}_2 \tilde {H} t_{R} \eta_3\eta_1^2\nonumber\\ 
&+&\frac{h^u_{31}}{\Lambda^2} \bar{Q}_3 \tilde {H} u_{R} \eta_3^{\dagger^2} \!+\! \frac{h^u_{32}}{\Lambda^3} \bar{Q}_3 \tilde {H} c_{R} \eta_3^\dagger\eta_2^\dagger\eta_1^{\dagger^2} \!+\!  h^u_{33} \bar{Q}_3 H t_R\,.~~~~ 
\label{up-L}
\eea%
After the spontaneous symmetry breaking, the up-type mass matrix takes the form as
\bea%
m_u&=&\langle H \rangle h^u_{33}\left(
      \begin{array}{ccc}
       a \lambda_C^6&  \lambda_C^4 & \lambda_C^{10}  \\
        \lambda_C^6 & \lambda_C^2   & \lambda_C^6 \\
        \lambda_C^4   & \lambda_C^8 &1 \\
      \end{array}
    \right). 
\eea%
where $a=\frac{h^u_{11}}{h^u_{33}}\sim{\cal O}(0.1)$. 
The left handed rotation required to diagonalize this matrix is given by, %
\bea%
V_L^u=\left(
\begin{array}{ccc}
 -0.998 & 0.048 & 0 \\
 0.048 & 0.998 & 0.0001 \\
 0 & -0.0001 & 1 \\
\end{array}
\right),\eea
with the corresponding eigenvalues  
\bea
M_u= {\rm{diag}} (\lambda_C^7, \lambda_C^2,~ 1)\langle H \rangle
h^u_{33}~{\rm{GeV}}.%
\eea%
Therefore with $h^u_{33}$ of order unity, we can explain the order of magnitude of the 
up quark masses \cite{pdg}.

Similarly, the down mass matrix can also be obtained from the following Lagrangian,
 \begin{eqnarray}
\!\!\!\!\!{\cal L}_d \!\!&\!\!=\!\!&\!\! \frac{h^d_{11}}{\Lambda^3} \bar{Q}_1 H d_{R} \eta_3^2\eta_1
\!+\!  \frac{h^d_{12}}{\Lambda^3} \bar{Q}_1 H s_{R} \eta_2 \eta_3^2
 \!+\! \frac{h^d_{13}}{\Lambda^4} \bar{Q}_1 H b_{R} \eta_1 \eta_2^\dagger \eta_3^2\nonumber\\
&+&\frac{h^d_{21}}{\Lambda^3} \bar{Q}_2 H d_{R} \eta_1 \eta_2 \eta_3\!+\!
\frac{h^d_{22}}{\Lambda^2} \bar{Q}_2 H s_{R} \eta_2^\dagger \eta_3 \!+\!
  \frac{h^d_{23}}{\Lambda^2} \bar{Q}_2 H b_{R} \eta_3\eta_1\nonumber\\&+&
  \frac{h^d_{31}}{\Lambda^2} \bar{Q}_3 H d_{R} \eta_2\eta_1^\dagger
\!+\!\frac{h^d_{32}}{\Lambda^2} \bar{Q}_3 H s_{R} \eta_1^{\dagger^2}\eta_2^\dagger \!+\!  \frac{h^d_{33}}{\Lambda}
\bar{Q}_3 H b_{R} \eta_1^\dagger. ~~~~
\label{down-L}
\end{eqnarray}
After the spontaneous symmetry breaking, the down mass matrix takes the form, 
\bea%
m_d&=&\langle H \rangle h^d_{33} \lambda_C^2\left(
      \begin{array}{ccc}
       \lambda_C^4&  \lambda_C^4 & \lambda_C^6  \\
        \lambda_C^4 & b\lambda_C^2   & \lambda_C^2 \\
        \lambda_C^2   & \lambda_C^4 &1 \\
      \end{array}
    \right),
    \eea%
where
$b=\frac{h^d_{22}}{h^d_{33}}\sim {\cal O}(0.2)$ is considered so as to get the 
eigenvalues $M_d= (\lambda_C^4, \lambda_C^3,~ 1)\lambda_C^2\langle H \rangle h^d_{33}$ GeV.
Here also, with $h^d_{33}$ of order one, the order of magnitude estimate of the 
down quark mass hierarchies \cite{pdg} can be explained. 
The left handed rotation of the matrix $m_d$ is found to be%
\bea%
V_L^d=
\left(
\begin{array}{ccc}
 0.963 & -0.267 & 0.00023 \\
 -0.267 & -0.962 & 0.048 \\
 0.012 & 0.047 & 0.998 \\
\end{array}
\right).
 \eea%
Thus, the $V_{CKM}$ is found out to be , 
\bea%
V_{CKM}=V_L^{u^\dagger} V_L^d=\left(
\begin{array}{ccc}
 -0.975 & 0.22 & 0.002 \\
 -0.22 & -0.974 & 0.048 \\
 0.012 & 0.046 & 0.998 \\
\end{array}
\right).\eea%
So it is clear that a close to correct $V_{CKM}$ \cite{pdg} can be obtained from our framework by tuning the 
coupling involved in Eq.(\ref{up-L}) and Eq.(\ref{down-L}).


In conclusion, we have developed a flavor symmetric approach to realize the neutrino masses and mixing through 
the inverse seesaw mechanism. The flavor symmetry consists of a non-abelian $\Delta$(27) group as well as two other 
discrete symmetries $Z_4$ and $Z_3$. We have also imposed a 2-3 symmetry on the Largangian (in the lepton sector), 
the breaking of which plays instrumental role in realizing the deviation from a TBM structure. The inverse seesaw 
mechanism is characterized by a small lepton number violating Majorana mass term $\mu$, while the effective light 
neutrino mass is $m_\nu \propto \mu$. Therefore in the $2-3$ symmetric limit, one obtained a zero neutrino mass.  
By breaking this 2-3 symmetry in the vacuum alignment of the flavon $\sigma$ which generates the $\mu$ term, 
we find that the neutrino mass is generated and deviation from the TBM mixing is achieved simultaneously. 
This is a new result from the 
point of view of showing that the $\mu$ mass term which violates the lepton number has the same origin as the 
deviation from the TBM limits. Within this model, the mass hierarchies in the charged lepton sector are also obtained.
Finally we are able to show that a mere addition of single flavon ($\eta_3$) along with those are already present in 
constructing the lepton sector can explain the up and down sector mass hierarchies as well as a close to the correct 
$V_{CKM}$ for some natural values of the parameters involved in the model. 

\section{Acknowledgment}
A.S acknowledges the hospitality of CFP, Zewail City of Science
and Technology, Cairo during a visit when this work was initiated.
S.K acknowledges partial support from the  European Union FP7  ITN INVISIBLES (Marie Curie Actions, PITN- GA-2011- 289442).
The work of M.A is partially supported by the ICTP grant AC-80.


%
\end{document}